# CCD PHOTOMETRY OF RR LYRAE STARS IN M5 AS A TEST FOR THE PULSATIONAL SCENARIO.

E.Brocato[1], V.Castellani[1,2], V.Ripepi[3]

[1]Osservatorio Astronomico Collurania, I-64100 Teramo, Italy
[2]Dipartimento di Fisica, University of Pisa, I-56100 Pisa, Italy
[3]Dipartimento di Fisica, Università della Calabria, I-87036 Cosenza, Italy

# Abstract

In this paper we present new CCD investigations of RR Lyrae pulsators in the Oo.I globular cluster M5. We confirm the variability of 11 Gerashchenko (1987) objects, adding the evidence for further 16 variables in the cluster central region. B V curves of light for 15 RR Lyrae are presented. With the addition of further 11 curves of light by Storm, Carney and Beck (1991) one is dealing with a sample of 26 well studied cluster pulsators whose properties have been implemented with similar data for RR Lyrae in clusters M3, M15, M68 to allow a comparison with the theoretical scenario recently presented by Bono and Stellingwerf (1994). On this basis, we discuss the distribution of stars in the period amplitude diagram, disclosing a substantial reduction of Sandage's period shift.

We suggest that theoretical constraints concerning periods and amplitudes could allow information on masses and luminosity of the pulsators directly from Bailey's diagram only. Static temperatures have been derived for all stars in the sample, discussing the dependence on the temperature of the observed pulsational properties.

# 1. Introduction

Since many years the pulsational properties of RR Lyrae stars in galactic globular clusters represent an exciting field of investigation, open to contrasting results and, in turn, to competitive interpretations. As an example, the intrinsec luminosity of these stars is still debated (see Buonanno, Corsi and Fusi Pecci, 1989, Sandage, 1990b, Caputo et al., 1993 and references therein), thus leaving room to alternative hypotheses about the origin of the well known separation of galactic globular clusters in two Oosterhoff groups. One may notice that investigations on that matter have been largely affected by the lack of firm theoretical constraints on the global phenomenology of pulsations, as due to the difficulties in the theoretical approach which have prevented an exhaustive exploration of the theoretical pulsational scenario.

However, such a theoretical situation is improving in recent time, thanks to the non linear, non local and time dependent convective models presented by Bono and Stellingwerf (1994, hereafter BS) which allows for the first time the pulsational amplitude to be related to the evolutionary parameters of pulsating stars. Similar progresses in the theoretical scenario obviously stimulates a corresponding improvement in the observational data, to produce a data base as reliable and as exhaustive as possible, to be compared to the growing amount of theoretical constraints.

According to such a scenario, we already presented (Brocato, Castellani and Ripepi, 1994) a CCD investigation of RR Lyrae stars in the globular M68, taken as an interesting but still unexplored representative of Oo.II type galactic globular clusters. In that paper we brought to the light pulsational evidences supporting the close similarity between M68 and the well studied cluster M15 when a reddening difference by $\Delta E(B-V) = 0.04$ is taken into account. In this paper we present new CCD BV photometry for RR Lyrae stars in the well know globular M5, enlarging the sample of RR Lyrae already presented by Storm, Carney and Beck (1991, hereafter SCB). The reason for this choice was primarily the study of a RR Lyrae rich Oo.I cluster where a large amount of variables still lack of accurate photometry, the original investigations having been performed by Oosterhoff (1941) only in photograpic V band.

In the next section we discuss observations and data reduction. As a result, we present new curves of light in both B and V bands for 15 cluster variables, deriving for all these objects static effective temperatures. In addition, we confirm the variability of 11 suspected variables given by Gerashchenko (1987), discovering further 16 variables in the cluster central region.

In section 3 data from M5 will be implemented with pulsational data for selected clusters in order to allow a comparison with the theoretical pulsational scenario. The analysis will be particularly devoted to a discussion of pulsational amplitudes, disclosing a relevant qualitative agreement with new theoretical predictions.

## 2. Observation and data reduction.

Photometric procedures have been already presented in a previous paper devoted to the cluster C-M diagram (Brocato, Castellani and Ripepi 1995: Paper I) and will be in the following only shortly recalled. Observations have been performed at the 1.5 m Danish ESO telescope at La Silla during the same run devoted to the study of M68, namely during the nights 14 to 17 April 1989. The chip was RCA 512x331, 0.5 arcsec/pixel sized. The cluster region was covered with four overlapping fields, as shown in Fig. 1 of Paper I. Exposure times ranged from 10 to 40 sec for V filter and from 20 to 90 sec for B, for a total of about 20 usable frames for each field and each color. Flat fields and bias exposures to correct CCD response to uniform sensitivity were obtained at the beginning and at the end of each night. According to the usual procedures data reduction was performed using ROMAFOT package for crowded photometry, reducing all the instrumental magnitudes to the instrumental magnitudes of our "best frames", namely frames 15 (B and V) of field 3. Instrumental magnitudes were finally calibrated by comparison with 88 stars in common with the CCD investigation by SCB with $V \leq 17$.

Analysis of the data confirmed the variability for all the objects but three (V25, V50, V51) in the Oosterhoff list. Taking advantage of the improved spatial separation in our CCD frames, the observational material allowed to check the variability of the 11 stars added by Gerashchenko (1987) to the Oosterhoff list as suspected variables, as well as to investigate the occurrence of further variables into the crowded cluster central region. To perform a similar search, we first estimated the dispersion of B-V colors among the 20 exposures, selecting as suspected variables the objects showing a dispersion larger than 0.02 mag., which have been further checked for variability. According to such a procedure we confirm the variability of all the 11 objects given by Gerashchenko (1987), discovering further 16 new variables in the crowded region, whose coordinates are given in Table 1.

Unfortunately, due to both the limited assigned observing time and the occurrence of a rather large amount of pulsators with periods around 0.5 days, we reached a sufficient sampling of the curves of light in both colors for only 15 RRs out of 52 already known in the field, other variables waiting for the allocation of further observing time. As an happy circumstance, we found that two out of our variables, V7 and V55 are in common with SCB and two, V59 and V79 in common with Cohen and Matthews, (1992). Fig. 1 compares our curves of light in both colors for V7 with SCB results, showing the agreement between these investigations. Comparison with data of Cohen and Matthews gives quite a similar agreement.

Data for our variables are presented in Table 2, where we report in the order: (1) the number of the variable, (2) the logarithm of the period, (3) the observed blue amplitude, (4,5) the mean V and (B-V) magnitudes, (6,7) the corrected (see section 4) and the dereddened (B-V) magnitudes, (8) log$T_e$ and (9) the bolometric magnitudes. Tables containing logs of observations for each variable are available in OACT Mosaic (Osservatorio Astronomico Collurania, Teramo; http://terri1.te.astro.it/oact-home/home.html). Fig. 2 gives the curves of light of all these variables in both B and V colors. Adding data the 11 variables presented by SCB, as given in the same Table 2, one is dealing with a sample of 26 curve of lights, which will allow the discussion given in the next section.

### 3. Pulsational amplitudes

Among the various observational parameters concerning variable stars, two parameters, namely the amplitude and the period, appear in principle of outstanding relevance, since their observational values can be easily derived and because they are unaffected by further observational or theoretical assumptions, like cluster reddening, distance modulus or color-temperature relations. This is of course the reason why since very early times Bailey (1899, 1902) arranged pulsational data in his well known "Bailey's diagram". Theoretical data presented by BS give us for the first time the opportunity of discussing observational data concerning the distribution in the Bailey diagram, as given in Table 2, in connection with theoretical prescriptions on the matter.

However, one has preliminarly to notice that BS computations only refer to a model by 0.65 M, Y=0.30, i.e., to a mass which could be (but could be not) appropriate for RR.s but for an amount of original He which is sensitively larger than the commonly adopted value Y=0.24 for Pop. II stars. As a result, the comparison can be regarded only as a first qualitative approach to the problem, firmer results requiring further information on the dependence of the theoretical results on the star mass and the original He content.

To approach the comparison, bolometric amplitudes given by BS have been translated in blue magnitudes adopting bolometric correction by Kurucz (1992). Of course this is a point where theoretical amplitudes become model (Kurucz) dependent; however one can estimate that possible errors should be of a minor relevance. On this basis we report in Fig. 3 theoretical expectations concerning the distribution in the Bailey's diagram for selected assumptions on the star luminosity. One has to notice the interesting occurrence that the expected distribution sensitively depends on the luminosity level, thus allowing - at least in principle- to read observational data in terms of star luminosities.

The same figure compare these theoretical prescriptions with the behaviour of variables in M5 as well as in well studied clusters membering both Oosterhoff type. To this purpose we selected photographic data presented by Sandage (1990a) for M3 and M15, together with the most recent large sample of CCD data presented by Walker (1994) for M68, all these data having been plotted in the upper panel of Fig. 3. Inspection of this figure reveals some relevant features worth to be discussed in some details. As a preliminary point one finds that data for M15 and M68 appear in good agreement. Recalling the very similar evolutionary status of stars in both clusters, one can conclude that, as for amplitudes, photographic data are as good as more recent CCD data. Suggesting, in turn, that the similarity between data for M3 and M5 is real, and not an artifact of the different observing procedure.

As a further point, one finds that whereas first overtone pulsators show the well known dichotomy marking the membership of an Oosterhoff group, fundamental pulsators do not show any clear separation in the Bailey plane, both distributions showing interesting qualitative features in common with theoretical predictions.

To look better in such a behavior, one can get rid of the variation in periods driven by variation in the star luminosity by adopting the so called "reduced periods", i.e., correcting the periods for the difference in luminosity, thus reducing the periods to the values expected for a common value of the star luminosities (Sandage 1981). However, the evaluation of reduced period deserves a preliminary discussion. As already quoted, the procedure gives the periods expected when all stars are reduced to a common luminosity level. The problem is the choice of this level. Following the procedure adopted by Sandage (1981) in his pioneering work on that matter, one can take as reference the mean bolometric magnitude of the pulsators. However, this appears a risky procedure, since it place the pulsators at a luminosity, above the ZAHB locus, which depends on the dispersion in luminosity of the stars.

For this reason, we choose to make reference to a luminosity value supported by clear and well defined theoretical constraints, as given by theoretical prescriptions about ZAHB luminosities. A philosophy already adopted by Sandage (1990a), but not in connection with the distributions in Bailey's diagram. Following Bencivenni et al. (1991) we will take as reference the luminosity of the ZAHB at logTe=3.83.

However, the advantage of referring to a firm theoretical constraint, as the ZAHB luminosity, is facing the problem of a reliable identification of a lower envelope for the pulsators luminosities. Fig. 4 shows our choices about this observational parameter, as based on theoretical constraints about the dependence of HB luminosity on the temperature (Castellani, Chieffi and Pulone, 1991) and on the estimate of the luminosity level at which a sudden increase in the observed luminosity function of pulsators appears. We regard such an estimate at least as a safe upper limit for the ZAHB luminosity. However, Fig. 4 shows that in the all cases a safe lower limit can be easily put only 0.04 m fainter. On can estimate a possible systematic error as given by $\Delta logPc = -0.336\Delta m_{bol}$, thus not larger than $\Delta logPc=0.01$, which appears a rather negligible value.

Reduced periods are reported in the lower panel of Fig. 3, which shows, as expected, that the spread of data is sensitively reduced. Interesting enough, one finds that the slope of fundamental pulsators appears in good concordance with theoretical prescriptions, whereas c-type first overtone pulsators show evidences for the bell-shaped distribution foreseen by theory. However, as a most important point one finds that the sequence of fundamental pulsators does not show the period shift invoked by Sandage (1981) to support his suggestion about the difference between Oo.I and Oo.II pulsators, i.e., $\Delta logP(A(B)=const)=$ 0.055. As a matter of the fact, best fitting the distributions concerning Oo.I and Oo.II type pulsators one finds $\Delta \log P = 0.02 \pm 0.04$, i.e., a much lower period shift, if any. Since we use, among other the same data used by Sandage, we compared the two procedures, concluding that the difference is mainly produced by adoption made in this paper of updated Kurucz bolometric corrections.

As for comparison with theory, the distribution of fundamental pulsators cannot be taken as a real indication of the luminosity, until reliable indications about the dependence of such a topology from the stellar mass and He content will be available. However, one may notice that first overtone pulsators appear separated by a gap similar to the one foreseen by theory if Oo.II type pulsators were more luminous than Oo.I pulsators by about $\Delta \text{Log} L \simeq 0.1$, as expected on theoretical grounds. Thus we are inclined to regard such separation as a signature of the pulsator luminosity. If this is true, the distribution of fundamental pulsators should be taken as an evidence that the mass is probably playing a role. On this basis, we suggest that an exhaustive knowledge of theoretical constraints concerning periods and amplitudes could possibly allows to derive masses and luminosity of the pulsators from Bailey's diagram only. An occurrence which should of course be of paramount relevance.

## 4. Pulsation and temperatures.

According to a common procedure, observed mean colors can be translated into stellar temperatures, allowing the investigation of the relation between pulsation and temperatures. To obtain similar data, mean colors as evaluated in the present or in the already quoted papers, have been first corrected for reddening and thus according to procedure given by Bono, Caputo and Stellingwerf (1995) to derive the true color of static models.

Temperatures have been finally derived adopting the color temperature relation given by Kurucz (1992). Corrected colors and temperatures are all reported in the previous Table 2.

As a first test of stellar pulsation theory, we report in Fig. 5 the observed distribution of the reduced period versus temperatures for pulsators in the two Oo.I type clusters M3 and M5, as obtained assuming for M3 and M5 the reddening values E(B-V)= 0.00 and 0.03 (Sandage, 1990a, Harris and Racine, 1979, Burstein, Faber and Gonzalez, 1986), respectively. The same figure shows the expected distribution of stars for selected assumptions about star luminosities and/or masses as given by the well known vanAlbada and Baker (1971) relation. One find a rather good agreement, which is in part related to the use of recent Kurucz relations, previous relations ( as, e.g., Buser and Kurucz, 1978) giving a less accurate agreement.

When comparing the distribution in Fig. 5 to the prescriptions of stellar evolution one finds that data for both clusters arrange along a mean line which is in agreement with theoretical expectations concerning ZAHB pulsators in Oo.I type cluster, namely logL = $1.65 \pm 0.02$, M= 0.65 (see, e.g., Castellani, Chieffi and Pulone 1991). As an important point, one finds that the dispersion around this mean line is of the order of what expected on the basis of observational errors in colors ($\sigma_{max} = 0.02$). Since no dispersion in masses is revealed, one can only estimate that this last dispersion cannot exceed about 0.05 solar masses, which, however, is well beyond theoretical expectations.

Figure 6 (upper panel) shows the same distribution in Fig. 5 but for the cluster M15 and M68, adopting respectively E(B-V)= 0.07 and 0.03, as discussed in Brocato, Castellani and Ripepi (1994). Again one finds a good agreement between theory and observation. Comparison between Fig. 5 and the upper panel of fig. 6 visualize a well known occurrence, i.e., under our reddening assumptions, the difference between Oo.I and Oo.II clusters can be accounted for by only a slightly increase in the luminosity to mass ratio as foreseen by canonical evolutionary theory. However, the actual reddening of both clusters is largely debated, with some Authors preferring reddening larger by about 0.04 (see, e.g., Walker 1994). As shown in the same Fig. 6 (lower panel), increasing the reddening of both cluster by $\Delta$ E(B-V)=0.04 would increase the temperature by about $\Delta logTe$=0.012, with much more severe requirement about the variation of luminosity to the mass ratio, which would no more compatible with the current evolutionary scenario.

As a final puzzling feature, we present in Fig. 7 both theoretical and observational data concerning the relation between pulsational amplitudes and stellar temperature. Bearing in mind the warning about the assumptions made in the computations (i.e. only one model for M = 0.65 $M_\odot$ and Y =0.30), one can finds again a qualitative agreement between theory and observations, but with not negligible quantitative disagreements affecting both fundamental and first overtone pulsators. As a whole, we feel that a deeper discussion should wait a theoretical scenario extended to more reliable input parameters and, in particular, to more reliable values for the helium abundance.

However, internal comparison among observational data show that, with our assumption on the reddenings for Oo.II clusters, the distribution of fundamental pulsators appears rather independent of the Oosterhoff type, with only some indications for larger amplitudes in few M5 fundamental pulsators. Concerning the Oosterhoff controversy (see in particu-lar Sandage, 1981, 1990b, and Caputo, 1988) thus one finds the interesting result that if one takes low reddenings for Oo.II clusters, as in Fig. 7, and only in that case Sandage's suggestion for a common relation between amplitude and temperature appears fairly supported. But, in the same time, the evidence for Sandage's period shift would desappear from logP, logTe data.

## 5. Conclusion.

In this paper we present new CCD photometric results concerning RR Lyrae pulsators in the Oo.I globular cluster M5. We confirm the variability of 11 objects indicated by Gerashchenko (1987), adding the evidence for further 16 variables in the cluster central region. We present B V curve of light for 15 RR Lyrae. With the addition of further 11 curve of light presented by Storm, Carney and Beck (1991) one has a sample of 26 well studied cluster pulsators whose properties have been compared with data from clusters M3, M15, M68. We find that Sandage's period shift in the Bailey period amplitude diagram, if real, appears smaller than previous evaluation. This appears as a further difficulty faced by the suggestion of non canonical luminosities in globular clusters Horizontal Branch stars, to be added to previous difficulties (see Castellani, Degl'Innocenti and Luridiana 1993) which should be taken into account before attempting any modification of present evolutionary scenario.

Pulsational results are discussed in the frame of the theoretical scenario recently presented by Bono and Stellingwerf (1994) which for the first time allows to attempt an interpretation of the distribution of stars in the period amplitude diagram. On this basis, we suggest that an exhaustive knowledge of theoretical constraints concerning periods and amplitudes could possibly allows to derive masses and luminosity of the pulsators from Bailey's diagram only. Static temperatures have been finally derived for all stars in the sample, further discussing the temperature dependence of the observed pulsational properties.

# References

Bailey S.J. 1899, ApJ 10, 255

Bailey S.J. 1902, Ann. Harward Obs. 38, 132

Bencivenni, D., Caputo, F., Manteiga, M. and Quarta, M. L. 1991, ApJ, 380, 484

Bono, G. and Stellingwerf, R. F. 1994, ApJS, 93, 233

Bono, G., Caputo, F. and Stellingwerf, R. F. 1995, in press

Brocato, E., Castellani, V. and Ripepi, V. 1995, AJ, in press (Paper I)

Brocato, E., Castellani V. and Ripepi, V. 1994, AJ, 107, 622

Buonanno, R., Corsi, C. E. and Fusi Pecci F. 1989, A&A, 216, 80

Burstein, D., Faber, S. M. and Gonzalez, J. J. 1986, AJ, 91, 1130

Buser R. and Kurucz R. L. 1978, A&A, 70, 555

Caputo, F. 1988, A&A, 189, 70

Caputo, F., De Rinaldis, A., Manteiga, M., Pulone, L. and Quarta, M. L. 1993, A&A, 276, 41

Castellani, V., Chieffi, A. and Pulone, L. 1991, ApJS, 76, 911

Castellani V., Degl'Innocenti S. and Luridiana V. 1993, A&A 272,442

Cohen, J. G. and Matthews, K. 1992, PASP, 104, 1205

Harris, W. E. and Racine, R. 1979, ARA&A, 17, 241

Gerashchenko, A. 1987, IBVS, n. 3044

Oosterhoff, P. T. 1941, An. Lei., 144, 10

Kurucz, R. L. 1992, IAU Symp. 149, "The stellar population of galaxies", ed. Barbuy and Renzini (Kluwer)

Sandage, A. 1981, ApJ, 248, 161

Sandage, A. 1990a, ApJ, 350, 603

Sandage, A. 1990b, ApJ, 350, 631

Storm, J., Carney, B. W. and Beck, J. A. 1992, PASP, 103 1264

van Albada, T. S. and Baker, N. 1971, ApJ, 169, 311

Walker, A. R. 1994, 108, 555

**Figure captions**

Fig. 1: Photometry of V7 by SCB and from the present investigation.

Fig. 2: B and V light curves for the 15 RRs of our sample. The solid line represents the adopted fitting to data.

Fig. 3: Upper panel: A(B) vs log P, for the selected samples of variables S(90) = Sandage 1990a, W94 = Walker 1994. Lines shows theoretical results by BS for the labelled values of luminosities.
Lower panel: As the upper panel but for reduced periods (see text)

Fig. 4: $m_b$ vs log Te for variables in M3, M5, M68 and M15. Solid lines rapresent the location of ZAHBs. Bolometric magnitudes of ZAHBs at logTe=3.83 are also labeled.

Fig. 5: Reduced periods vs LogTe for variables in M5 and M3; solid lines represent the relation by van Albada and Baker (1971) for M=0.65 and for the labelled values of luminosities. Dashed lines show the same relation but for M=0.75.

Fig. 6: As in Fig. 5, but for M68 and M15. Upper panel $E(B - V)_{M68} = 0.03$ and $E(B - V)_{M15} = 0.07$. Lower panel: $E(B - V)_{M68} = 0.07$ and $E(B - V)_{M15} = 0.11$ (see text).

Fig. 7: A(B) vs log Te for the observational sample compared to theoretical prescriptions. Symbols and lines are as in Fig. 3. Arrows show the shift in Oo II temperature increasing the reddening by $\Delta E(B - V) = 0.04$

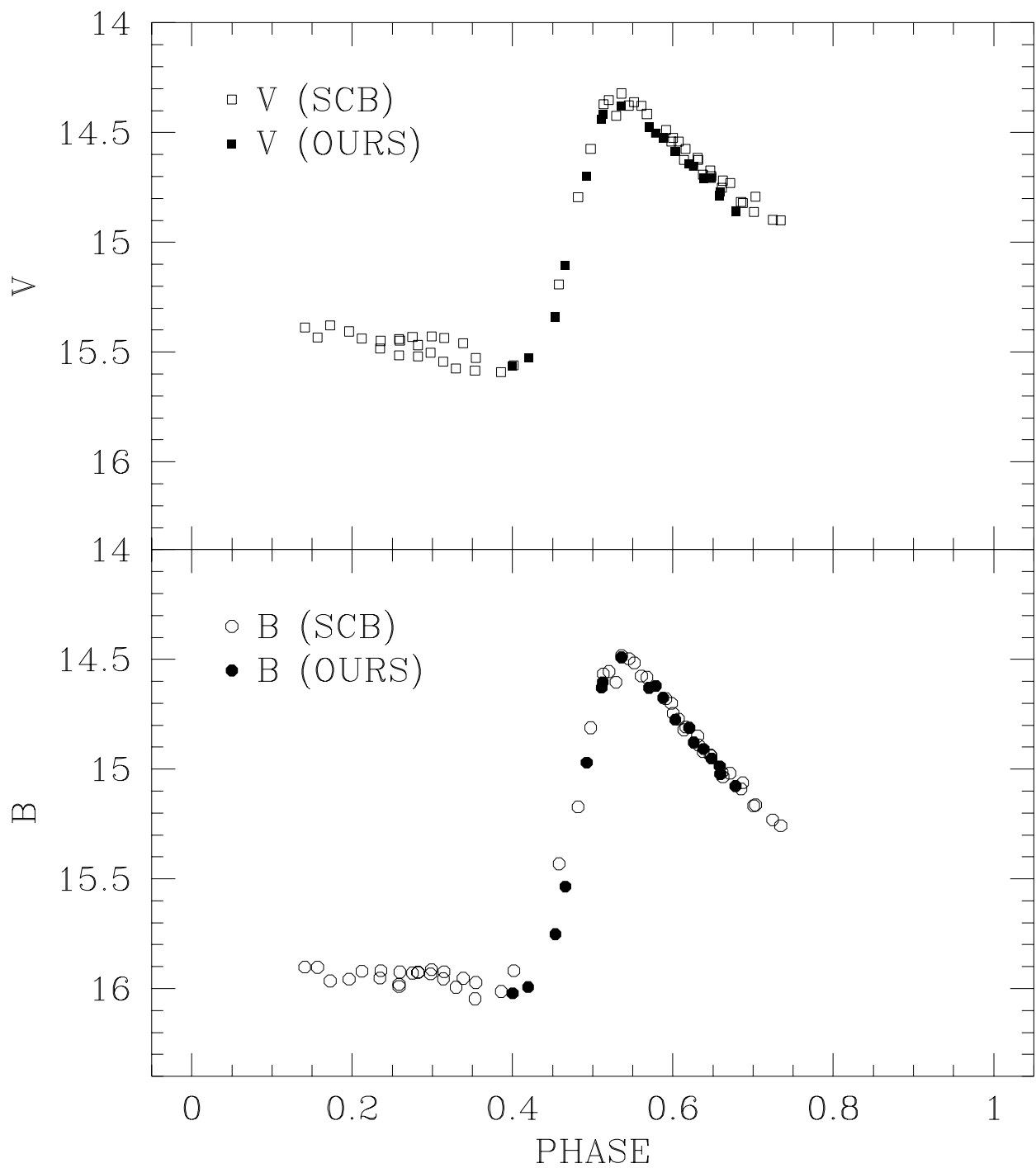

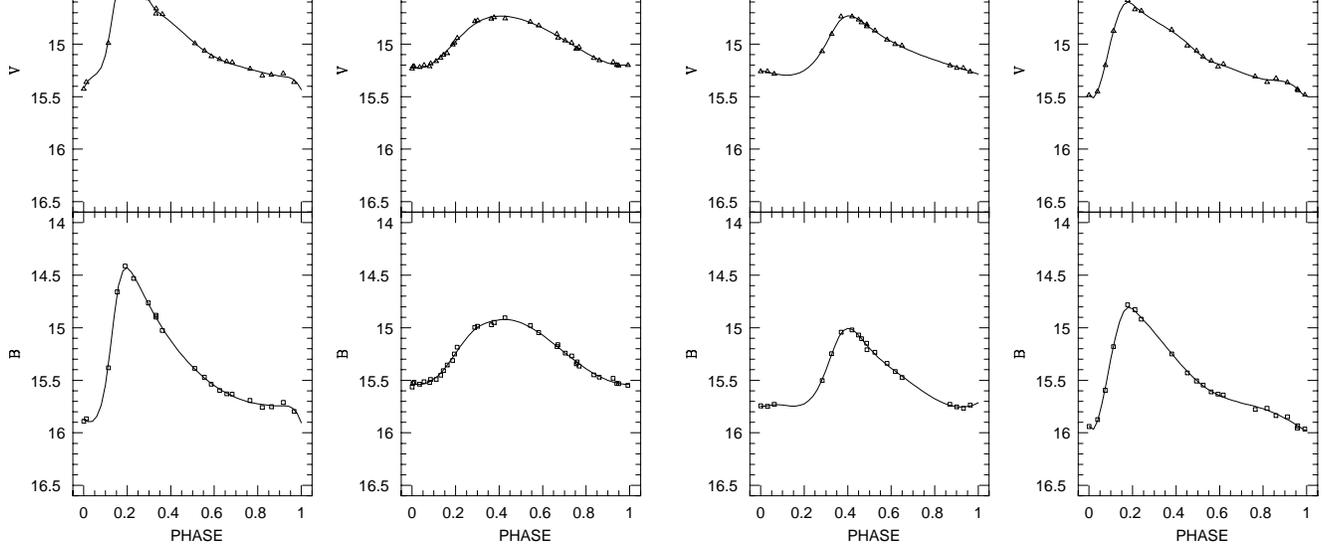
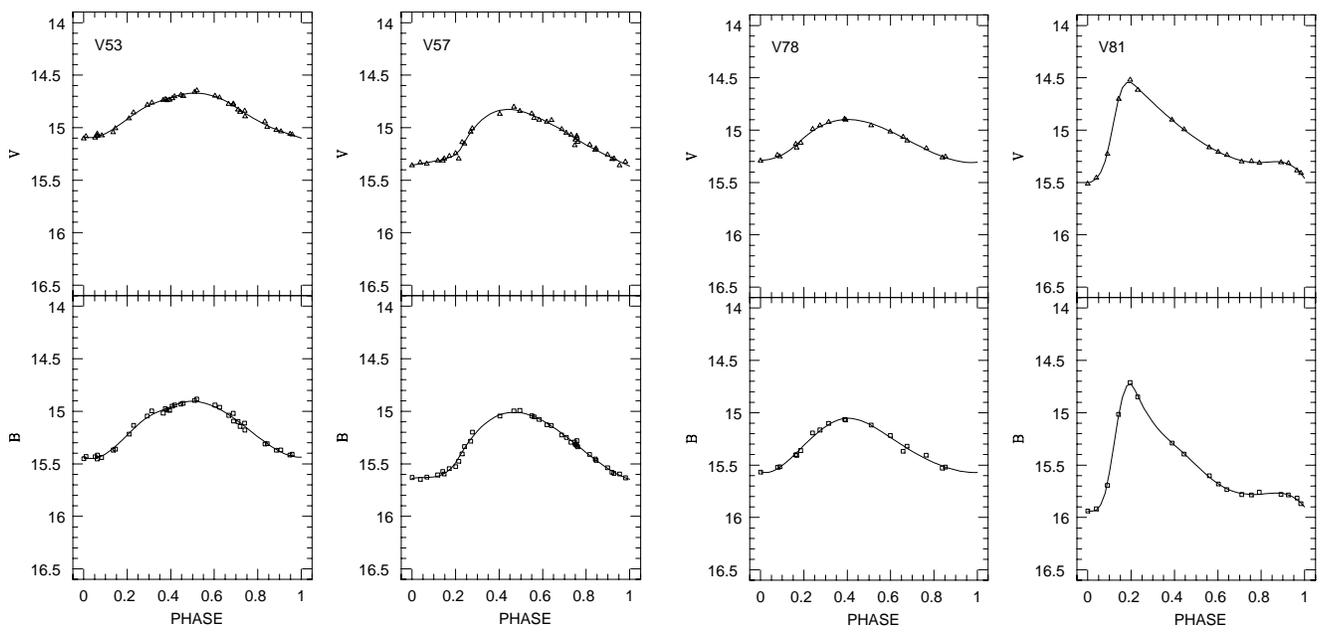
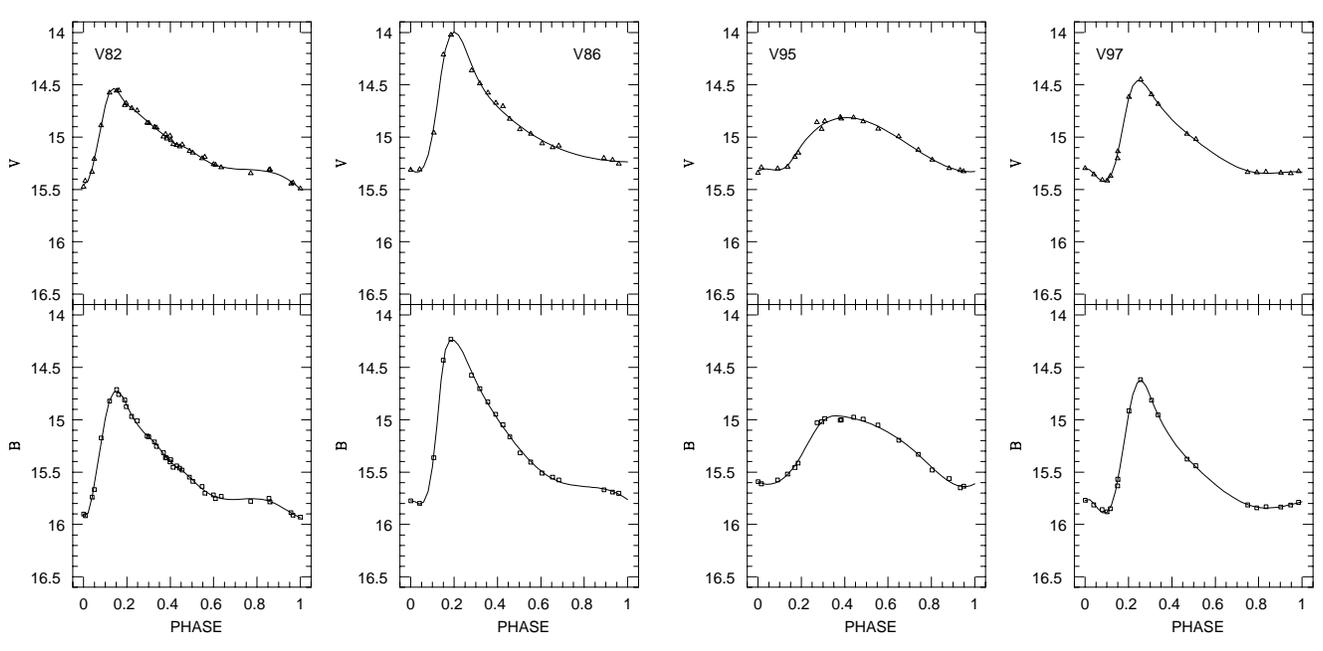

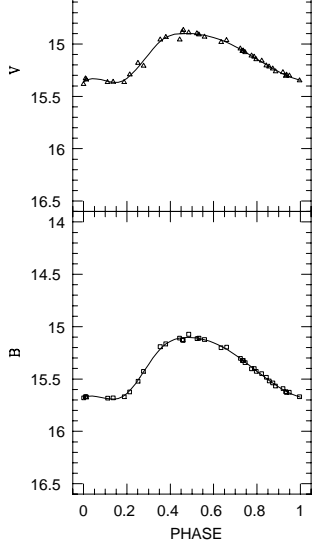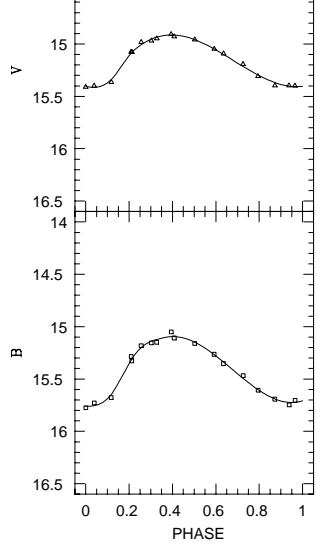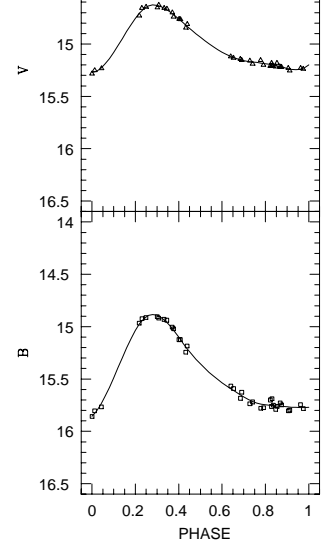

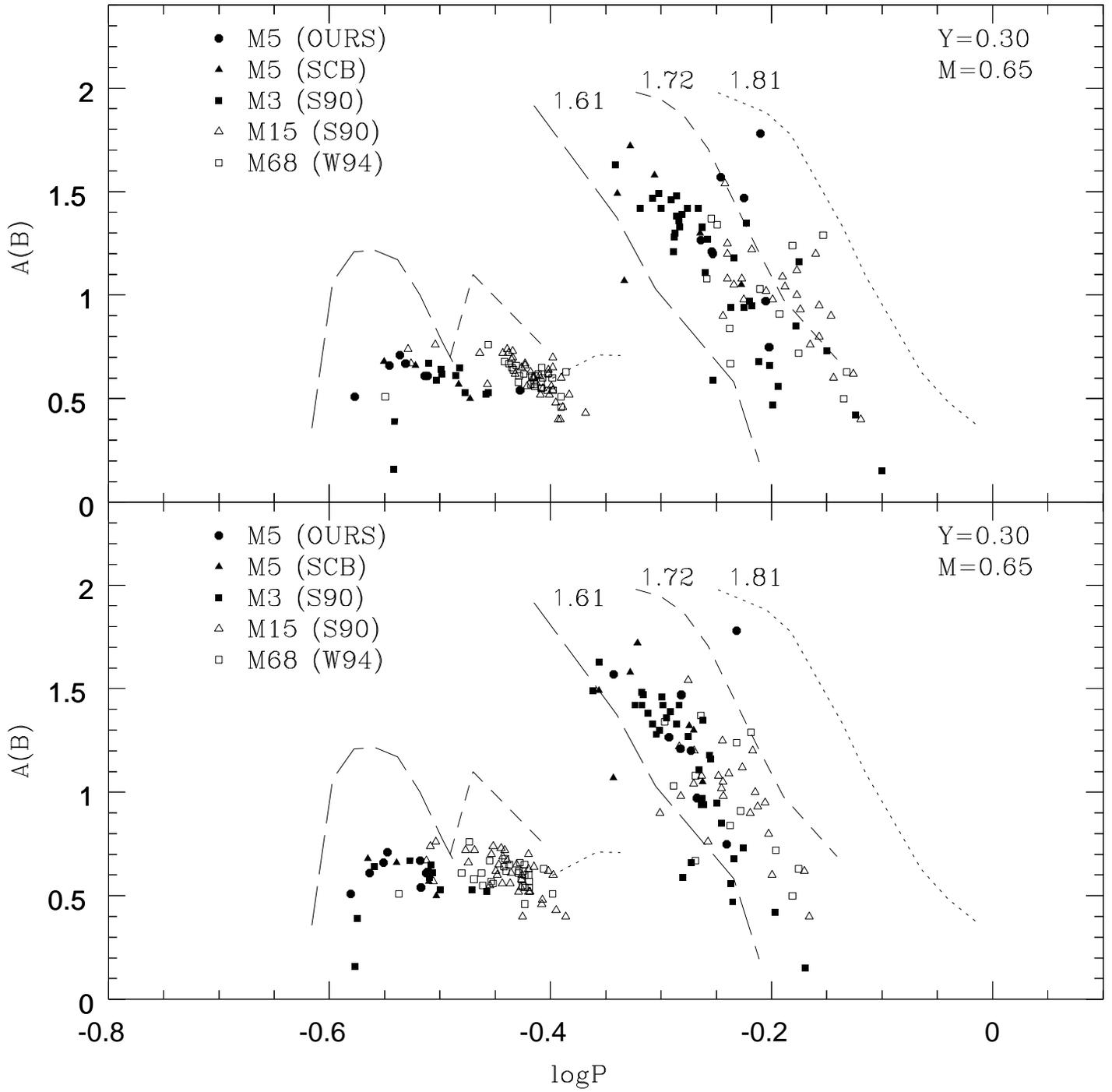

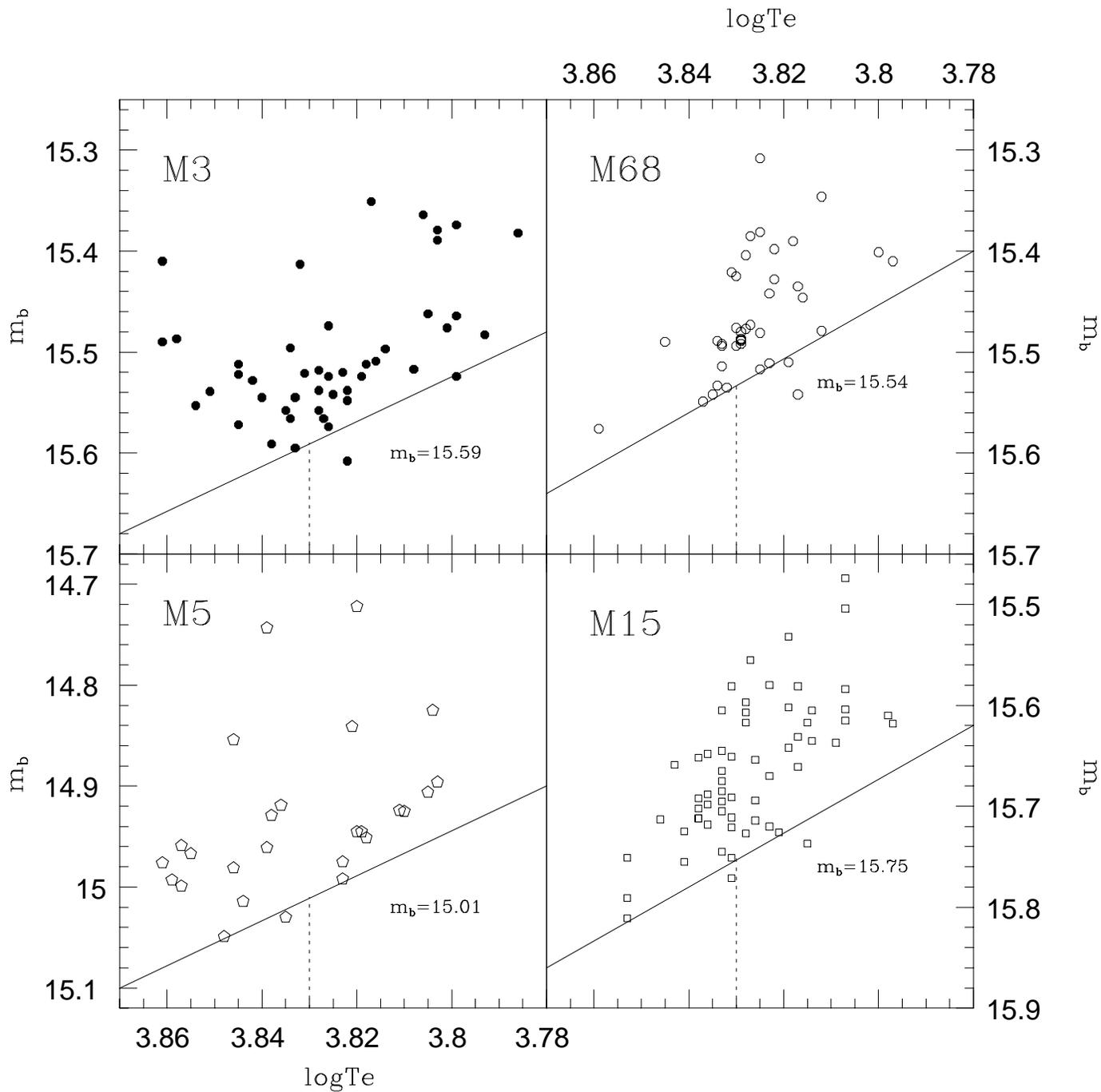

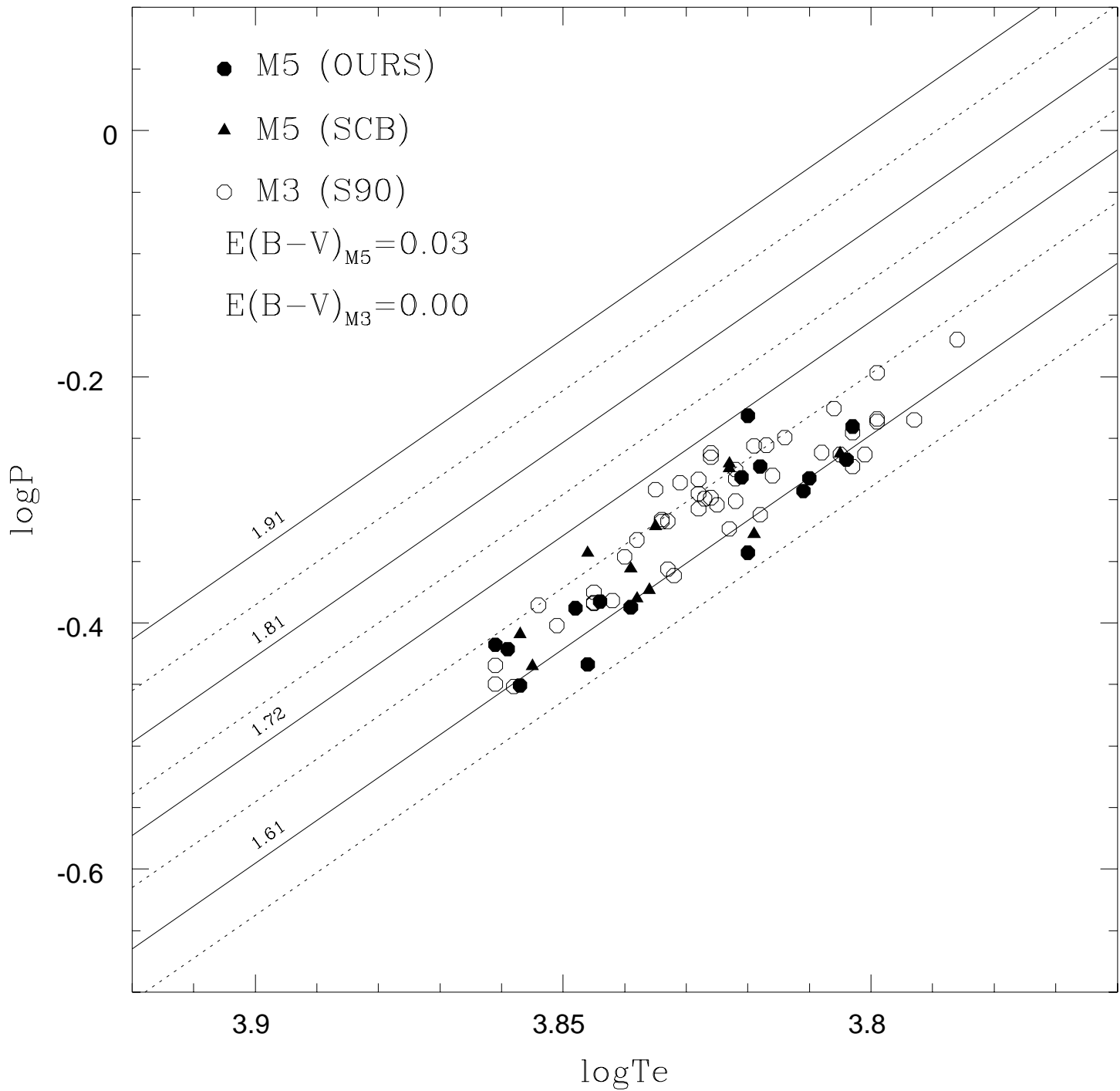

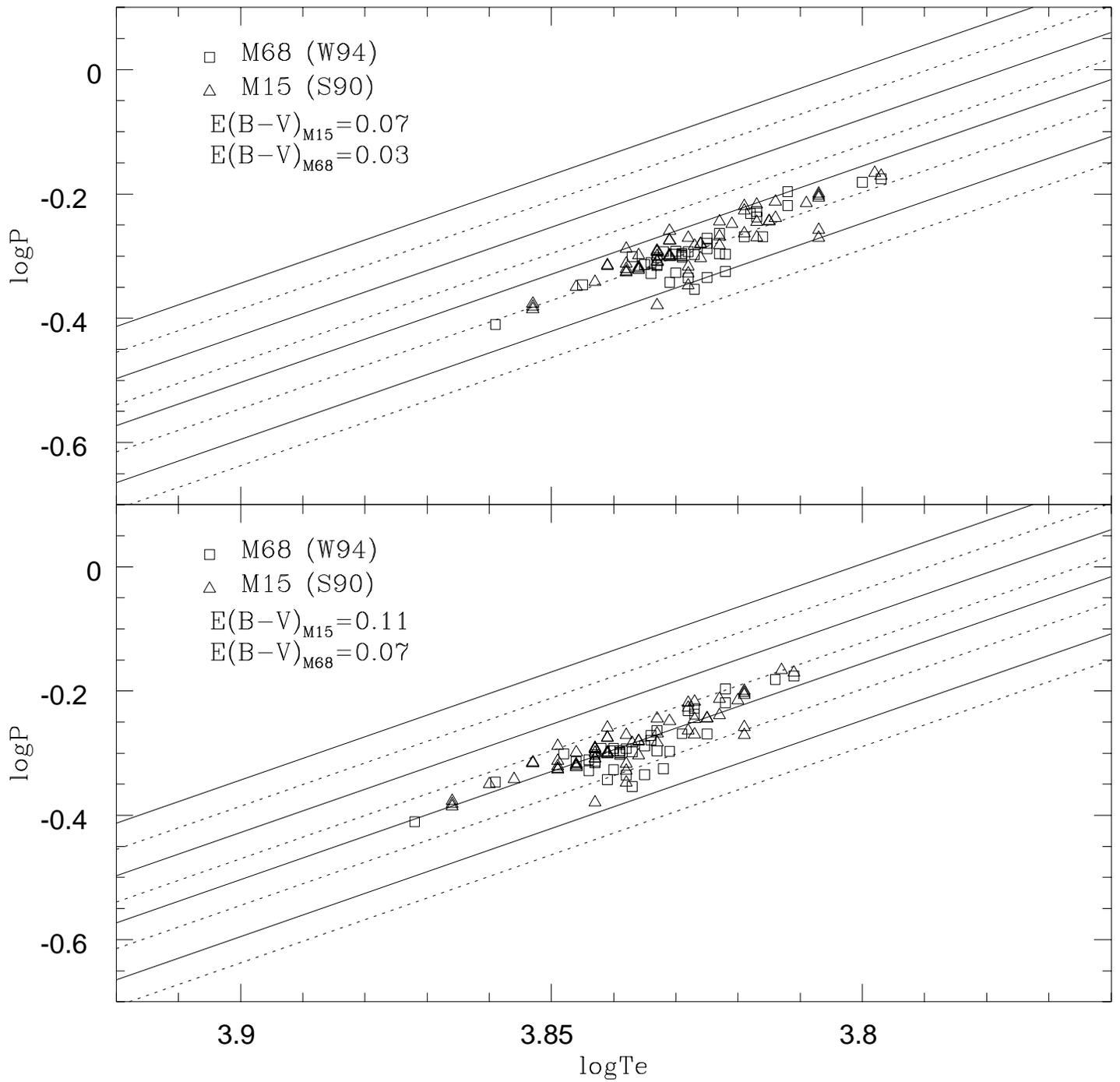

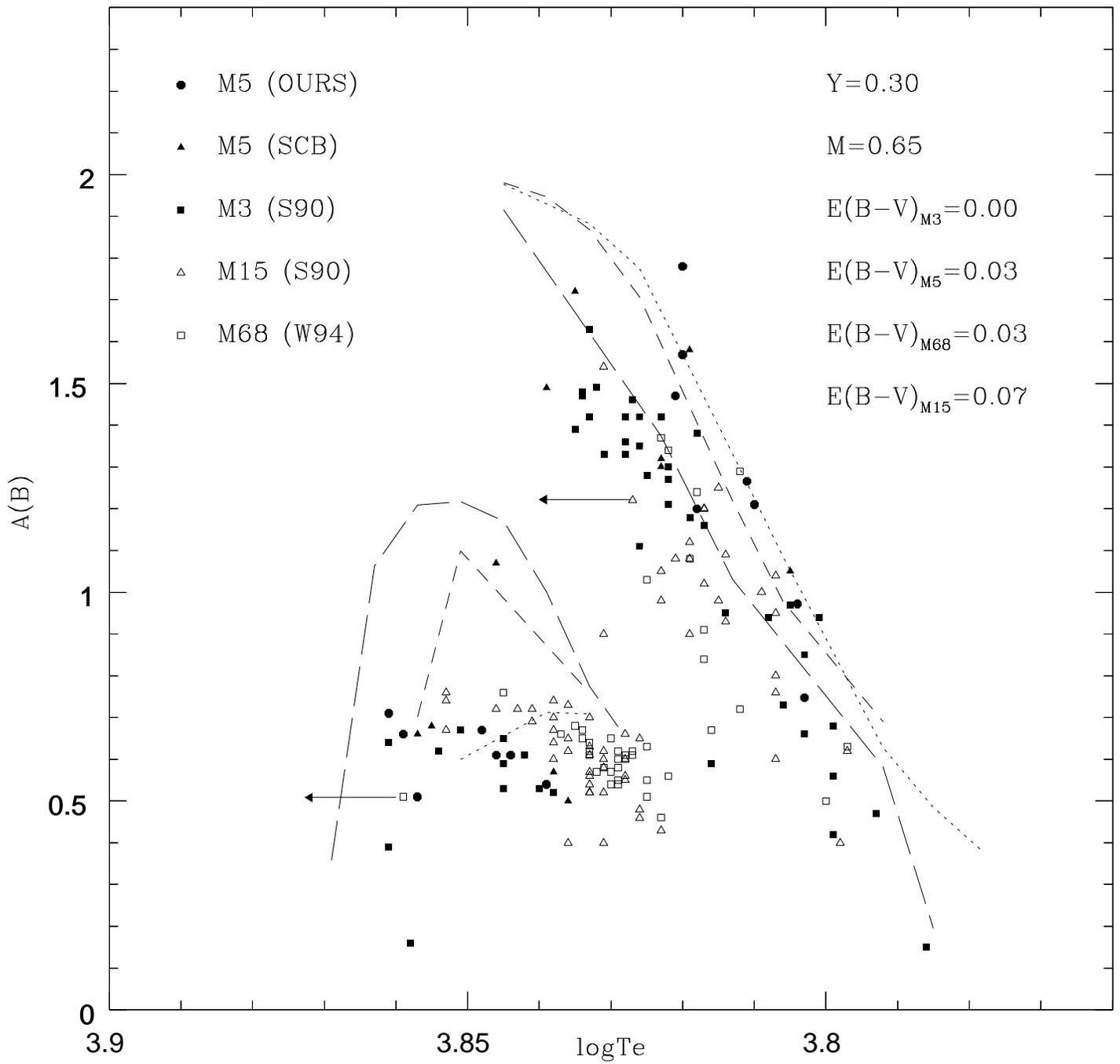

Tab1: New variables in M5. Coodinate are given with respect to the cluster center determined in paper I.

```
_______________________
  VAR       X         Y
         (arcsec)  (arcsec)
------------------------
  V115    -88.9    +52.9
  V116    -57.2     +1.0
  V117    -53.9     +0.2
  V118    -53.9    +36.6
  V119    -53.8     -7.6
  V120    -32.9    -52.7
  V121    -31.8     -1.2
  V122    -31.5    +12.4
  V123    -26.9     +3.8
  V124    -25.2     -3.3
  V125    -23.6    -19.7
  V126    -15.1     +4.3
  V127     -6.3    +26.3
  V128     -1.3    -10.9
  V129    +10.6    +13.9
  V130    +31.7    -11.2
```

Tab2: Relevant quantities for RR Lyrae in M5 obtained in the present work and by SCB (see text).

| Var | logP | A(B) | V | (B-V) | (B-V)c | (B-V)oc | (logTe)oc | mbol |
|-----|------|------|---|-------|--------|---------|-----------|------|
| (1) | (2)  | (3)  | (4) | (5) | (6)    | (7)     | (8)       | (9)  |
| 11  | -0.2248 | 1.470 | 14.996 | 0.378 | 0.365 | 0.335 | 3.821 | 14.841 |
| 35  | -0.5113 | 0.610 | 14.972 | 0.262 | 0.258 | 0.228 | 3.846 | 14.854 |
| 36  | -0.2022 | 0.748 | 15.079 | 0.436 | 0.426 | 0.396 | 3.803 | 14.896 |
| 45  | -0.2100 | 1.780 | 15.102 | 0.392 | 0.369 | 0.339 | 3.820 | 14.945 |
| 53  | -0.4276 | 0.540 | 14.872 | 0.292 | 0.290 | 0.260 | 3.839 | 14.743 |
| 57  | -0.5456 | 0.660 | 15.098 | 0.228 | 0.219 | 0.189 | 3.859 | 14.993 |
| 78  | -0.5771 | 0.510 | 15.106 | 0.226 | 0.226 | 0.196 | 3.857 | 14.999 |
| 81  | -0.2539 | 1.210 | 15.096 | 0.413 | 0.403 | 0.373 | 3.810 | 14.925 |
| 82  | -0.2530 | 1.200 | 15.111 | 0.387 | 0.377 | 0.347 | 3.818 | 14.951 |
| 86  | -0.2462 | 1.569 | 14.879 | 0.385 | 0.370 | 0.340 | 3.820 | 14.722 |
| 95  | -0.5364 | 0.710 | 15.079 | 0.220 | 0.211 | 0.181 | 3.861 | 14.976 |
| 97  | -0.2639 | 1.266 | 15.093 | 0.410 | 0.400 | 0.370 | 3.811 | 14.924 |
| 98  | -0.5137 | 0.610 | 15.134 | 0.269 | 0.264 | 0.234 | 3.844 | 15.014 |
| 100 | -0.5312 | 0.670 | 15.165 | 0.258 | 0.251 | 0.221 | 3.848 | 15.049 |
| 122 | -0.2053 | 0.972 | 15.006 | 0.432 | 0.422 | 0.392 | 3.804 | 14.825 |
| SCB | | | | | | | | |
| 7   | -0.3059 | 1.580 | 15.103 | 0.388 | 0.372 | 0.342 | 3.819 | 14.945 |
| 8   | -0.2626 | 1.320 | 15.128 | 0.369 | 0.359 | 0.329 | 3.823 | 14.975 |
| 15  | -0.4727 | 0.500 | 15.052 | 0.303 | 0.303 | 0.273 | 3.836 | 14.919 |
| 18  | -0.3334 | 1.070 | 15.099 | 0.267 | 0.257 | 0.227 | 3.846 | 14.981 |
| 19  | -0.3279 | 1.720 | 15.165 | 0.330 | 0.310 | 0.280 | 3.835 | 15.030 |
| 28  | -0.2645 | 1.300 | 15.144 | 0.367 | 0.357 | 0.327 | 3.823 | 14.992 |
| 30  | -0.2276 | 1.050 | 15.086 | 0.430 | 0.420 | 0.390 | 3.805 | 14.906 |
| 31  | -0.5220 | 0.660 | 15.066 | 0.232 | 0.225 | 0.195 | 3.857 | 14.959 |
| 32  | -0.3393 | 1.490 | 15.089 | 0.303 | 0.289 | 0.259 | 3.839 | 14.961 |
| 55  | -0.4829 | 0.570 | 15.059 | 0.299 | 0.294 | 0.264 | 3.838 | 14.929 |
| 62  | -0.5507 | 0.680 | 15.076 | 0.239 | 0.231 | 0.201 | 3.855 | 14.967 |